\documentclass[epjST]{svjour}
\usepackage{graphics,epsfig}

\newcommand{\bi}[1]{\bibitem{#1}}

\newcommand{\ba}{\begin{eqnarray}}
\newcommand{\ea}{\end{eqnarray}}
\newcommand{\beqs}{\begin{eqnarray}}
\newcommand{\eeqs}{\end{eqnarray}}

\begin{document}
\title{Generalized hadron structure and elastic scattering
}
\author{O.\,V. Selyugin\fnmsep\thanks{\email{selugin@theor.jinr.ru}  }
}
\institute{
JINR,  Bogoliubov Laboratory of Theoretical Physics, 
141980 Dubna,  Russia }
\abstract{The new parameterization of the Generalized Parton Distributions (GPDs) 
t-dependence is investigated. It is shown that the new form of the GPDS allows one 
to reproduce sufficiently well the electromagnetic form factors of  the proton and 
neutron at small and large momentum transfer and obtain  a good description of 
the elastic nucleon-nucleon scattering at high energies.  
  }
\maketitle
\section{Introduction}

  The new experimental data of the LHC, especially of the TOTEM Collaborations 
\cite{TOTEM1,TOTEM2}, require deep understanding of hadron interactions at high 
energies which are related with inner structure of the hadrons \cite{Rev-LHC}.
  The electromagnetic current of a nucleon is
\begin{eqnarray}
 J_{\mu} (P^{'},s^{'}; P,s)  = \bar{u} \left(P^{'},s^{'}\right) \Lambda_{\mu} (q,P) u(P,s)  \nonumber \\
 =  \bar{u}\left(P^{'},s^{'}\right) \left(\gamma_{\mu}
 F_{1}(q^2)\right)+\frac{1}{2M} i \sigma_{\mu \nu }q_{\nu }F_{2}(q^{2})u(P,s)\,,
\end{eqnarray}
where $P,s,  (P^{'},s^{'}) $ are the four-momentum and polarization
of the incoming (outgoing) nucleon
and $q = P^{'}- P $ is the momentum transfer.
The quantity  $  \Lambda_{\mu} (q,P) $ is the nucleon-photon vertex.

The Sachs form factors \cite{Ernst60} are related with the Dirac and Pauli form factors
   \ba
 G^{p}_{E}(t) = F^{p}_{1}(t) + \frac{t}{4M^2} F^{p}_{2}(t)\,;\qquad\qquad
 G^{p}_{M}(t) = F^{p}_{1}(t)  + F^{p}_{2}(t)\,;
\ea
 with 
  $(t = -q^2 < 0)$.
Their three-dimensional Fourier transform  provides the electric-charge-density
 and the magnetic-current-density distribution \cite{Sachs62}. Normalization 
requires $G^{p}_{E}(0) =1$,  $G^{n}_{E}(0) =0$  corresponding to proton and 
neutron electric charges; $G_{M} (0) = (G_{E} (0) +k) = \mu$ defines the proton 
and neutron magnetic moments. Here  $\mu_p = (1 + 1.79) \frac{e}{2M}$ is the 
proton magnetic moment and $k= F_{2}(0)$ is the anomalous magnetic moment: 
$k_p = 1.79$.

  The experiments based on the Rosenbluth separation method \cite{Rosenbluth},
  \ba
\frac{d \sigma}{d \Omega}=\frac{\sigma_{\rm Mott}}{\epsilon (1+\tau)}
 \left[\tau G^2_{M}(t) + \epsilon G^2_{E}(t)\right]
 \ea
 where  $\tau= Q^2/4M^2_p$ and $\epsilon = [1+2(1+\tau) \tan^2(\theta_e/2)]^{-1}$
 is a measure of the virtual photon polarization,  suggested that  the scaling behavior 
of both the proton form factors 
  approximately are described by the dipole form
\ba
G^{p}_{E} \approx \frac{G^{p}_{M}}{\mu_{p}} \approx \frac{G^{n}_{M}}{\mu_{n}}
\approx G_{D} = \frac{\Lambda^4}{(\Lambda^{2} -t)^2}\,,
\ea
which leads to
\ba
 F_{1}^{D} (t) = \frac{4M_{p}^{2} - t \mu_{p}}{4M_{p}^{2} - t } \ G_{D}\,; \qquad\qquad
F_{2}^{D} (t) =  \frac{4 k_{p} M_{p}^{2}}{ 4 M^{2}_{p} - t} \ G_{D}\,;
\ea
with $\Lambda^2= 0.71$\,GeV$^2$.\,

 The information about the form-factors can also been obtained  
 by using the polarization method \cite{Akhiezer,Arnold}.
 Using the longitudinal components of the recoil proton polarization in the electron
 scattering plane,  the ratio of the form factors
 \ba
 \frac{G^{p}_{E} }{ G^{p}_{M} } = -\frac{ P_t }{ P_l } \frac{ E+E^{'}}{2 M_{p} } \tan(\theta/2)
 \ea
 can be obtained. These data manifested a strong deviation from the
 scaling law and disagreement with the data obtained by the Rosenbluth technique.
 The results consist in   an almost linear decrease of $G^{p}_{E} / G^{p}_{M}$.

The electromagnetic  form factors can be obtained from the first moments
of the Generalized Parton Distributions (GPDs) \cite{DMuller94,Ji97,R97,Collins97}
\ba
 F_{1}(t) = \sum_{q} \ \int^{1}_{0} \ dx  \ {\cal{ H}}^{q} (x, t)\,;\qquad\qquad
 F_{2}(t) =\sum_{q} \  \int^{1}_{0} \ dx \  {\cal{E}}^{q} (x,  t)\,,
\ea
following from the sum rules \cite{Ji97,R97}.

  The Regge-like picture for GPDs  was proposed in \cite{R04}
\ba
{\cal{H}}^{q} (x,t) \  \approx  \frac{1}{x^{\alpha'  \ (1-x) \ t}} \ q(x)\,.
\ea

    In \cite{Yuan03,Burk04}, it was shown that at large $x  \rightarrow 1$
 the behavior of GPDs requires a  power of $(1-x)$  
\ba
{\cal{H}}^{q} (x,t) \approx {\rm exp}\left[ a \ (1-x)^n\,t\right] \ q(x)\,.
\ea
with $n \geq 2$. It was noted that $n=2$ naturally leads to the Drell--Yan--West 
duality between parton distributions at large $x$ and the form factors.

 \section{ Transfer dependence of General parton distributions }

 In \cite{ST-PRD09}, the new form of the momentum transfer dependence of 
the General Parton Distribution was proposed
\ba
{\cal{H}}^{q} (x,t) \  = q(x) \  {\rm exp} \left[  a_{+}  \ t \
(1-x)^2/x^{m}\right ]\,.
\ea
  The value of the parameter $m=0.4$ is fixed by the low $t$  experimental data 
while the free parameters $a_{\pm}$ ($a_{+} $ =-- for ${\cal{H}}$ and $a_{-} $ 
-- for ${\cal{E}}$) were chosen to reproduce the experimental data. A large 
momentum transfer behavior corresponds to $x \approx 1$ in~(10), where the 
dependence on $m$ is weak. For the parton distribution we choose the PDFs 
obtained in  \cite{MRST02}.

Correspondingly, we obtain
 \ba {\cal{E}}^{q} (x,t)  =
{\cal{E}}^{q} (x) \,\exp \left[  a_{-}  \ t \ (1-x)^{2}/x^{0.4}\right]\,.
\ea
 with
 \ba
{\cal{E}}^{u} (x)   = \frac{k_u}{N_u} (1-x)^{\kappa_1} \ u(x)\,,   \qquad\qquad
{\cal{E}}^{d} (x)   = \frac{k_d}{N_d} (1-x)^{\kappa_2} \ d(x)\,,
\ea
 where $\kappa_1 =1.53$ and $\kappa_2=0.31$  \cite{R04}.

According to  the normalization of the Sachs form factors $k_u=1.673\,, 
\quad  k_d=-2.033\,,   \quad   N_u=1.53\,, \quad  N_d=0.946$.

 \section{Proton and neutron electromagnetic form factors }
The ratio of the Suchs proton form factors is shown in Fig. 1. Our 
theoretical curve coincides with the polarization data. \vspace{3mm}

\begin{figure}[!h] 
~\vglue -1.5cm
\centerline{
\includegraphics[width=.6\textwidth]{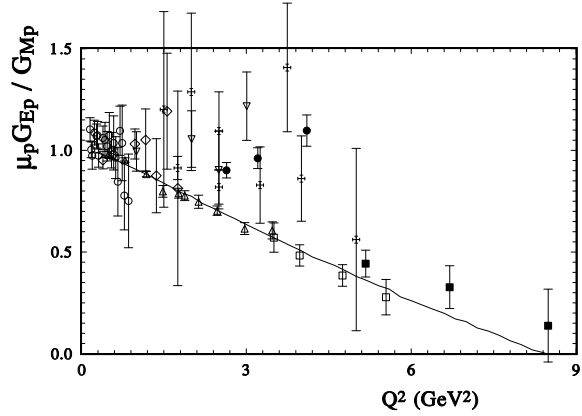}
}
 \caption{ $\mu_{p} G_{E}^{p}/G_{M}^{p}$ (the line corresponds
 to model calculation; the experimental data are
from  \cite{Gayou01,Gayou02,Punjabi05,Arr05,Hu06,Graw07,Qat05,Qat06}).
 } \label{Fig_3}
\end{figure} \vspace{-3mm}

   It is to be  noted, that some changing of the size of $a_{-}$ can 
lead to the change of the slope of the ratio $\mu_{p} G_{E}^{p}/G_{M}^{p}$ 
and leads to the Rosenbluch data. However, the neutron data on the 
momentum dependence of the form factors do not allow this.

  The isotopic invariance can be used to relate the proton and neutron GPDs.
 Hence, we do not change any parameter and keep the same $t$-dependence 
of GPDs as in the case of proton. As a result, we obtain the good description 
of the electromagnetic form factors of the neutron too.

\section{Gravitational form factors of the nucleon}

Taking the matrix elements of the energy-momentum tensor $T_{\mu\nu}$ \cite{Ji97}
 \ba
 \left\langle p^{\prime}|\hat{T}^{Q,G}_{\mu \nu} (0)|p\right\rangle &=& \bar{u}(p^{\prime}) \biggl[
 A^{Q,G}(t) \frac{\gamma_{mu}P_{\nu} }{2}       +
 B^{Q,G}(t)\frac{i\left(P_{\mu}\sigma_{\nu \rho}+ P_{\nu} \sigma_{\mu \rho}\right) \Delta^{\rho} }{4 M_N}  \\ \nonumber
 &+& \ C^{Q.G}(t) \frac{  \Delta_{\mu} \Delta_{\nu} - g_{\mu \nu} \Delta^{2} }{M_N} \pm \ \bar{c}(t) g_{\mu \nu}
\biggr ] {u}(p)
  \ea
  one can obtain the gravitational form  factors of quarks which are related to the second  moments of GPDs.
 For $\xi=0$ one has
\ba
A_{q}(t) = \ \sum_q \ \int^{1}_{0} dx \ x {\cal{H}}_q(x,t); \,\  B_{q}(t)\ =  \ \sum_q \  \int^{1}_{0}  dx \ x {\cal{E}}_q (x,t).
\ea

This representation combined with our model allows one to calculate  the 
gravitational form factors of valence quarks and their contribution (being just 
their sum) to the gravitational form factors of the nucleon.  Our results for 
$A_{u+d}(t)$ are shown in  Fig.~2 a (for small momentum transfer) and in 
Fig.~2 b (for large momentum transfer). We can describe the obtained form 
of $A_{Gr.}(t)$ by the standard dipole form but with  a sufficiently large
  size of  $\Lambda^2= 1.8$ -- 2.0\,GeV$^2$. The corresponding results 
are shown in Fig.~2 a,b.

\begin{figure}[!h ] 
~\vglue -1.5cm
\includegraphics[width=.5\textwidth]{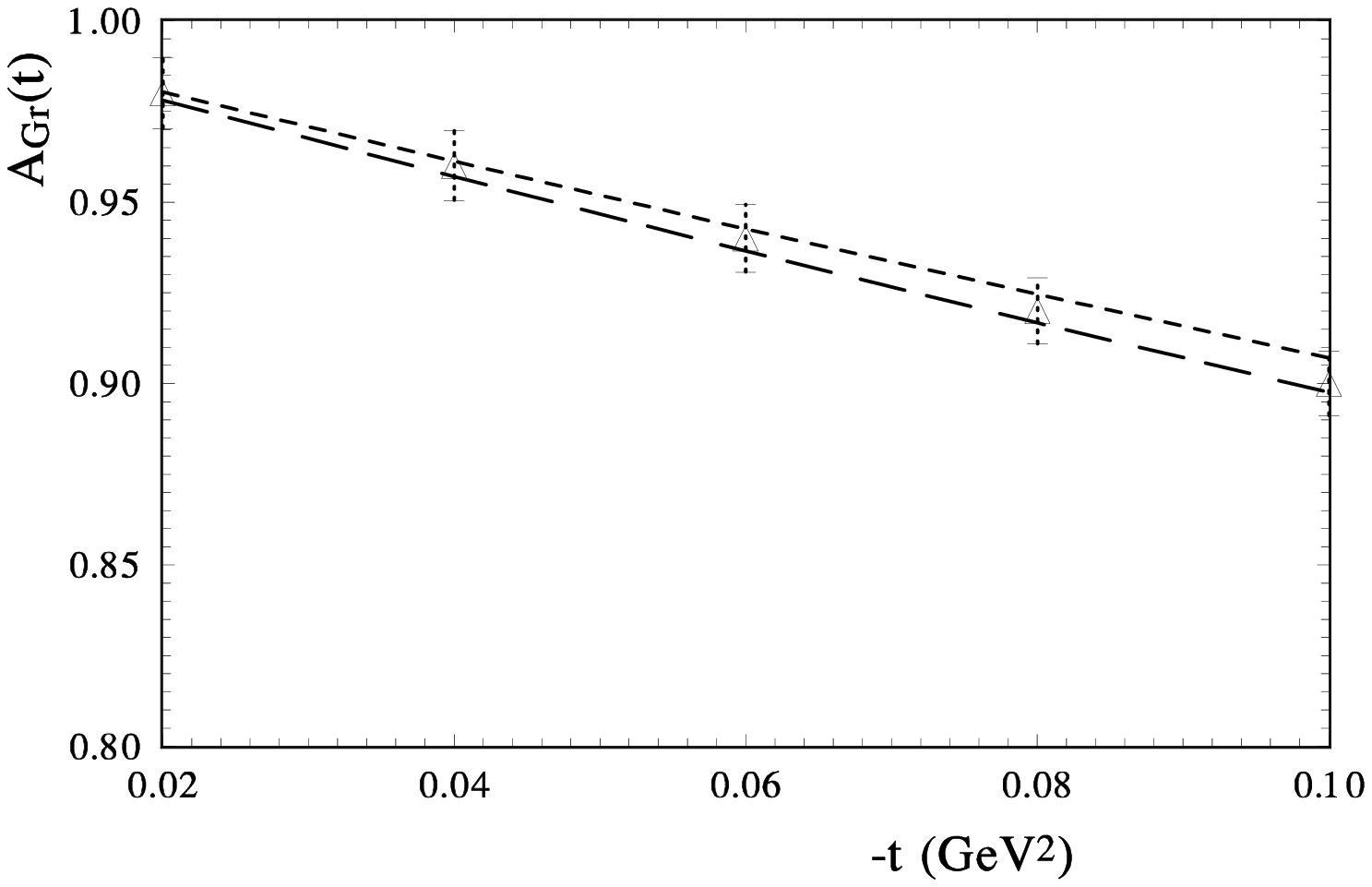}
\includegraphics[width=.5\textwidth]{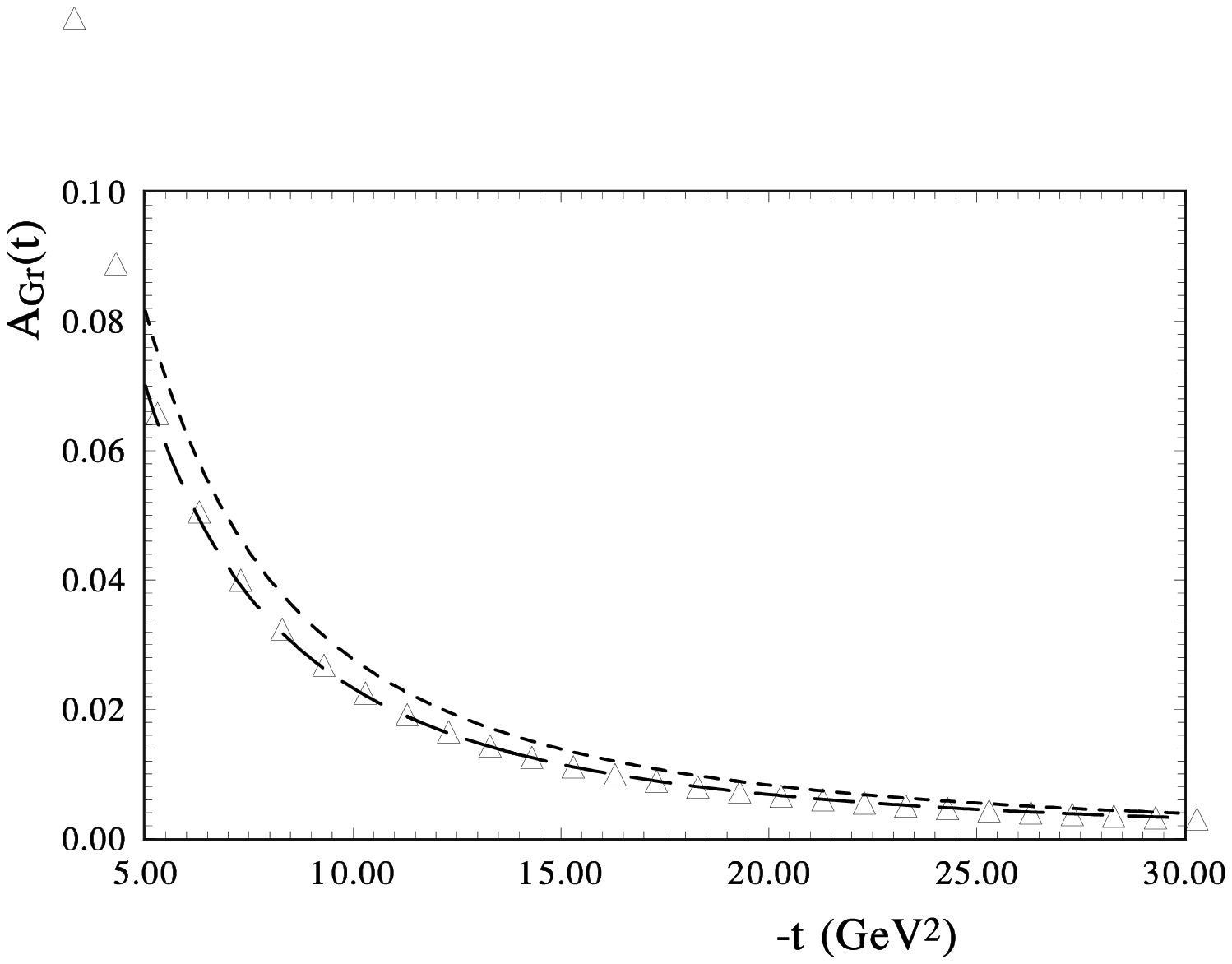}
 \caption{ $A_{Gr.}{t}$ (hard and dot-dashed lines correspond to the variant with 
$\Lambda^2=2.0$\,GeV$^2$ and  $\Lambda^2=1.8$\,GeV$^2$; triangles are 
the integrals of the second momentum of the GPDs).
 } \label{Fig_2}
\end{figure}

   Of course, the dipole presentation of $A_{gr.}(t)$ is the simplest form close  to 
the standard form of the electromagnetic form factors. A slightly more complicated 
form can be taken as
\ba
A_{Gr.}(t) = \ 1/\left[1+q/\Lambda_1 - t/\Lambda_2\right]^2,
\ea
where $q^2=-t$ and $\Lambda_1$ and $\Lambda_2$ are determined by fitting 
the obtained value of $A_{Gr.}(t)$. However, firstly it is necessary to examine 
the different PDF sets obtained by the different collaborations from the analysis 
of the dip inelastic scattering data to see which PDFs  better describe the 
electromagnetic form factors.

\section{\boldmath The $pp$ and $p\bar{p}$ elastic differential cross sections 
at high energies}

  The differential cross sections of the nucleon-nucleon elastic scattering  can be 
written as the sum of the different spiral scattering amplitudes: Every amplitude 
$\phi_{i}(s,t)$, including the electromagnetic and hadronic forces,  can be 
expressed as
\begin{eqnarray}
  \phi(s,t) =
  F_{C} \exp{(i \alpha \varphi (s,t))} + F_{N}(s,t) ,
\end{eqnarray}
with
 $ \varphi(s,t) =  \varphi(t)_{C} - \varphi(s,t)_{CN} $,
 where   $\varphi(t)_{C}$ appears in the second Born approximation of
 the pure Coulomb amplitude, and the term $\varphi_{CN}$ is
 defined by the Coulomb-hadron interference   \cite{Sel-ph1,Sel-ph2,Sel-phsum}.

 In \cite{Sel-M1}, it was proposed that the hadron scattering amplitude  includes 
two parts which are proportional to two forms of the form factors. One form factor 
is related to the charge distribution to the hadron and  represented in the standard 
form of the electromagnetic form factor - $G(t)$. The second form factor is proportional
 to the momentum energy tensor distribution  and can be represented by the 
gravitational form factor -$A(t)$. Both the forms of the form factors can be obtained 
from the same GPDs as first and second moments.

      The hadron Born term of the  elastic nucleon amplitude can be written as
  \begin{eqnarray}
 F_{N}^{\rm Born}(s,t)  =  h_1 \ G^{2}(t) \, F_{1}(s,t) \, \left(1+R_1/\hat{s}^{0.5}\right) 
                  +   h_{2} \  A^{2}(t) \ F_{2}(s,t) \left (1+R_2/\hat{s}^{0.5}\right)
\end{eqnarray}
  where $F_{a}(s,t)$ and $F_{b}(s,t)$  has the standard Regge form 
  \begin{eqnarray}
 F_{1}(s,t)  = \hat{s}^{\epsilon_1} \ e^{B(s) \ t}\,; \qquad\qquad
 F_{2}(s,t)  = \hat{s}^{\epsilon_1} \ e^{B(s)/4 \ t}\,,
\end{eqnarray}
 with $G(t)$ being the  electromagnetic form factor relative to the first moment of GPDs 
and $A(t)$ relative to the second moment of GPDs.
\begin{eqnarray}
G(t) =\frac{L_{1}^{4}}{\left(L_{1}^2-t\right)^2} \ \frac{4m_{p}^{2}- (1+k) \ t}{4m_{p}^{2}- \ t}\,;  \qquad
A(t)= \frac{L_{2}^4}{\left(L_{2}^2-t\right)^2}\,;
 \label{overlap}
 \end{eqnarray}
with the parameters:  $L^{2}_{1}=0.71 $\,GeV$^2$; $L_{2}^2=2 $\,GeV$^2 $. 
The values of $R_1$ and $R_2$ are taken as free parameters
which reflect the rest of the low energy contributions to the high energy parts.
  Using the crossing symmetry properties of the scattering amplitudes
  of the proton-proton and proton-antiproton scattering at high energies
 we take  $   \hat{s}=s \ e^{-i \pi/2}/s_{0} $;  with $ s_{0}=1 $\,GeV$^2$.
 In this case, the real part of the scattering amplitude has no any additional 
parameters. The slope of the scattering amplitude has the standard logarithmic 
dependence on the energy.  $  B(s) = \alpha^{\prime} \ ln(\hat{s})$
  with $\alpha^{\prime}=0.24$\,GeV$^{-2}$.

\label{sec:figures} 
\begin{figure}[h] \vspace{-2cm}
\begin{center}
\includegraphics[width=0.8\textwidth] {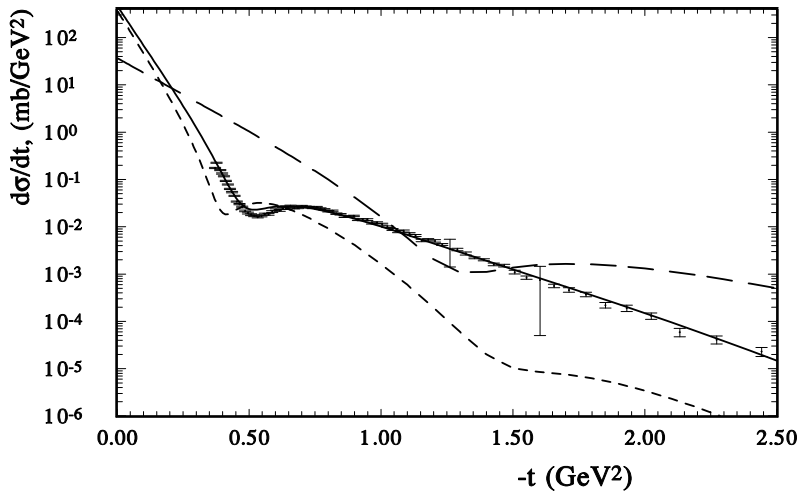}
\end{center}
 \caption{The differential cross sections of the $pp$ elastic scattering at  TeV. The hard line is the model
 predictions; the short dashed  and long dashed lines present the differential cross section in the case if we
 take into account the part of the scattering amplitude with the electromagnetic form factor or the gravitational form factors.
 } \label{Fig_2}
\end{figure} \vspace{-3mm}

  In the model,  the Born terms of the scattering amplitude  are calculated only. 
  Then the final scattering amplitude is obtained through the standard eikonalization 
procedure \cite{CPS-EPJ08,CS-PRD09}.  First, we calculate the eikonal phase in the 
impact parameter representation and then, after integrating over the impact parameter,
 we obtain the final  scattering amplitude. In Fig. 3, the parts of the differential scattering
 corresponding to the Born parts of the scattering amplitude are shown.  Of course, the 
final result is not the simple sum of the separate parts determined by the separate Born 
terms. Only the eikonalization of the Born amplitude leads to the good representation of 
the experimental differential cross sections. In Fig.~3, we can see that the first term
 of the scattering amplitude, proportional to the electromagnetic form factor, determines 
the differential cross sections at small momentum transfer. At large momentum transfer 
the second term of the Born amplitude, proportional to the gravitational form factors, 
gives the dominant  contribution.

As a result, we can obtain a good description of the existing differential cross sections
  of the proton-proton and proton-antiproton at high energies \cite{Sel-M1}, and the 
model prediction coincides with the new TOTEM data at 7\,TeV. In Fig.~3, the predictions 
of the model for the 7\,TeV are shown. Except the size of the diffraction minimum
   the coincidence of the model and experimental data is very good.

\section{Conclusion }
 The new data of the TOTEM Collaboration \cite{TOTEM1,TOTEM2} presented the 
behavior of the differential cross sections at small and sufficiently large momentum 
transfer (up to $-t=2.5$\,GeV$^2$. These results do not coincide with the predictions 
of the most popular models of hadron interactions at  high energies. We show that 
the using  two form factors - the electromagnetic and the gravitational form factors,
 which are obtained from the same General Parton Distributions as first and second 
moments respectively, we can obtain the description of all high energy elastic 
scattering data in small and large momentum transfer regions. This opens the new 
way for connecting the deep inelastic scattering processes, from which the parton 
distributions were obtained, with the elastic hadron scattering.


 {\it The author would like to thank J.-R. Cudell  for helpful discussion. 
   }

\end{document}